\documentclass {revtex4}
\usepackage {amsmath}
\usepackage {amstext}
\usepackage {latexsym}
\usepackage {xspace}
\input {epsf}
\begin{document}
%\draft
%\preprint{HEP/123-qed}
% --- /home/prasad/.lib/tex/vgarticle.sty ---
% =====================================================================
% Personal article style			v guruprasad,sep1999
% =====================================================================

% ---------------------------------------------------------------------
% General favourite definitions

% Orderly dates 
\newcommand\DateYMD{
	\renewcommand\today{\number\year.\number\month.\number\day}
	}

\newcommand\DateDMY{
	\renewcommand\today{\number\day.\number\month.\number\year}
	}

\newcommand\DateDmmmY{
	\renewcommand\today{
		\number\day\space
		\ifcase\month\or
		Jan\or Feb\or Mar\or Apr\or May\or Jun\or
		Jul\or Aug\or Sep\or Oct\or Nov\or Dec\fi
		\number\year
		}
	}

% Tables:
\newenvironment{mptbl}{\begin{center}}{\end{center}}
\newenvironment{minipagetbl}[1]
	{\begin{center}\begin{minipage}{#1}
		\renewcommand{\footnoterule}{} \begin{mptbl}}%
	{\vspace{-.1in} \end{mptbl} \end{minipage} \end{center}}

% Figures:
\newif\iffigavailable
	\def\figavailable{\figavailabletrue}
	\def\nofigavailable{\figavailablefalse}
	\figavailable% default

\newcommand{\Fig}[4][bh]{
	\begin {figure} [#1]
		\centering\leavevmode
		\iffigavailable\epsfbox {\figdir /#2.eps}\fi
		\caption {{#3}}
		\label {f:#4}
	\end {figure}
}

% Deflists:
\newlength{\defitemindent} \setlength{\defitemindent}{.25in}
\newcommand{\deflabel}[1]{\hspace{\defitemindent}\bf #1\hfill}
\newenvironment{deflist}[1]%
	{\begin{list}{}
		{\itemsep=10pt \parsep=5pt \topsep=0pt \parskip=10pt
		\settowidth{\labelwidth}{\hspace{\defitemindent}\bf #1}%
		\setlength{\leftmargin}{\labelwidth}%
		\addtolength{\leftmargin}{\labelsep}%
		\renewcommand{\makelabel}{\deflabel}}}%
	{\end{list}}%

% Equation numbering:
\makeatletter
	\newcommand{\numbereqbysec}{
		\@addtoreset{equation}{section}
		\def\theequation{\thesection.\arabic{equation}}
		}
\makeatother

% ---------------------------------------------------------------------
% General settings

\DateYMD			% the only rational default
\def\figdir{.}		% redefine this locally

% =====================================================================

% --- /home/prasad/.lib/tex/vgabbr.sty ---
% =====================================================================
% Personal abbreviations			v guruprasad,sep1999
% =====================================================================
% Scientific:

\def\arcdeg{\hbox{$^\circ$}}
\newcommand{\bold}[1]{\mathbf{#1}}
\newcommand\degree{$^\circ$}
\newcommand{\Qed}{$\bold{\Box}$}
\newcommand{\order}[1]{\times 10^{#1}}
\newcommand{\Label}[1]{\ \\ \textbf{#1}}
\newcommand{\Prob}[1]{\mathrm{\mathbf{Pr}}[#1]}
\newcommand{\Expect}[1]{\mathrm{\mathbf{E}}[#1]}

% Latinora:

\newcommand\ala{\emph{a la}\xspace}
\newcommand\vs{\emph{vs.}\xspace}
\newcommand\enroute{\emph{en route}\xspace}
\newcommand\insitu{\emph{in situ}\xspace}
\newcommand\viceversa{\emph{vice versa}\xspace}
\newcommand\terrafirma{\emph{terra firma}\xspace}
\newcommand\perse{\emph{per se}\xspace}
\newcommand\adhoc{\emph{ad hoc}\xspace}
\newcommand\defacto{\emph{de facto}\xspace}
\newcommand\apriori{\emph{a priori}\xspace}
\newcommand\Apriori{\emph{A priori}\xspace}
\newcommand\aposteriori{\emph{a posteriori}\xspace}
\newcommand\nonsequitor{\emph{non sequitor}\xspace}
\newcommand\visavis{\emph{vis a vis}\xspace}
\newcommand\primafacie{\emph{prima facie}\xspace}

% http://www.liv.ac.uk/education/hd/latin.html
\newcommand\circa{\emph{c.}\xspace}
\newcommand\ibid{\emph{ibid.}\xspace}		% previous citation
\newcommand\loccit{\emph{loc.\ cit.}\xspace}	% cited in the ref
\newcommand\opcit{\emph{op.\ cit.}\xspace}	% cited in the ref
\newcommand\viz{viz{}\xspace}			% videlicet - NO STOP

% http://www.tsolv.com/schools/lghs/clubs/latin/Latin_Abbreviations.html
\newcommand\ie{i.e.{}\xspace}			% id est
\newcommand\eg{e.g.{}\xspace}			% exempli gratia
\newcommand\etal{\emph{et al.}\xspace}		% et alii, et alibi
\newcommand\cf{cf.{}\xspace}			% confer (compare)
\newcommand\etc{etc.{}\xspace}			% et cetera

\newcommand{\LD}{\begin{description}}
\newcommand{\DE}{\end{description}}
\newcommand{\LI}{\begin{itemize}}
\newcommand{\LE}{\end{itemize}}
\newcommand{\LN}{\begin{enumerate}}
\newcommand{\NE}{\end{enumerate}}
\newcommand{\VB}{\begin{verbatim}}
\newcommand{\VE}{\end{verbatim}\\}
\newcommand{\QB}{\begin {quotation}}
\newcommand{\QE}{\end {quotation}}

\newcommand{\Or}{\vee}
\newcommand{\Def}{\stackrel{\triangle}{=}}
\newcommand\s{s$^{-1}$\xspace}
\newcommand\ssq{s$^{-2}$\xspace}
\newcommand\bra[1]{\langle#1|}
\newcommand\ket[1]{|#1\rangle}
\newcommand\iprod[2]{\langle #1 | #2 \rangle}
\newcommand\tuple[2]{\langle #1, #2 \rangle}
\newcommand\up{\uparrow}
\newcommand\dn{\downarrow}
\renewcommand{\thefootnote}{\fnsymbol{footnote}}

% =====================================================================

% --- epsf.tex ---
% --- title.tex ---
\title {
	Contraction and distension by tidal stress and
	its role as the cause of the Hubble redshift
	}
\author {V Guruprasad}
\email {prasad@watson.ibm.com}
\affiliation {
	IBM T J Watson Research Center,
	Yorktown Heights, NY 10598, USA.
	}
\DateYMD
%\date{1999.10.6}		% transferred to prd
%\date{1999.9.17}		% first communication from aps
\begin {abstract}
% --- abs.tex ---
I show that
a cumulative contraction or expansion must result from
repetitive tidal action in a curved stress field,
depending on the direction of the curvature.
The resulting expansion of solid materials onboard deep space probes
and the corresponding contraction on earth
would be of the right magnitude to account for
all aspects of the Pioneer anomaly,
leading to the two component model previously proposed.
Importantly,
I show via signal path analysis that
the anomaly mathematically implies planetary Hubble flow, and that
it is predicted by the standard model equations
when the cosmological constant is also taken into account
at this range.
Also shown is that
the variations of the anomaly do not permit a different explanation.
The prediction of the Hubble flow occurring as a result
in the view of the shrinking observer is now fully explained
from both quantum and Doppler perspectives,
fundamentally challenging the cosmological ideas of the past century.
Lastly, I discuss how
the contraction reconciles the geological evidence
of a past expansion of the earth.
\end {abstract}
% --- body.tex ---
\maketitle
%\clearpage
\section {Introduction}
\label {s:intro}

I describe below
a hitherto unsuspected form of plastic flow in solids
affecting the very dimensions of the lattice,
\viz an extremely slow but unavoidable contraction or expansion
under the combined action of
repetitive tidal stress and a nonuniform field of force.
This combination of stresses is new to science
as the curvature of the earth's surface was never considered
as a factor in prior studies of the solid state.
I further show that the phenomenon should translate
to all scales of the internal structure of matter.
The effect itself,
in the instances to be described, is extremely small,
of $10^{-18}$~\s or less,
corresponding to a ``half-life''
matching the age of the solar system.

Our negligence of it as in the past is no longer justifiable, however,
as even this small effect is sufficient to produce
a very pronounced consequence,
as already predicted separately
\cite {Prasad2000c}:
that of making incoming light appear redshifted
in exact proportion to the distance $r$ of its source from us, \ie
of making the stars appear to be receding at velocities $v = Hr$
conforming to Hubble's law, and
to be accelerating in their recession,
according to the \emph{same} law,
at rates $\ddot{r} \equiv dv/dt = d(Hr)/dt = H dr/dt = H^2 r$,
as $H$ would be inherently independent of $r$
for the present mechanism.
The positive cosmological constant $\Lambda$
discovered in 1998 from observations of
Type Ia supernovae (SNe Ia)
happens to be exactly of this magnitude ($H^2$)
\cite {Reiss1998}
\cite {Garnavich1998a}
\cite {Leibundgut1998}
\cite {Garnavich1998b},
as is the anomalous acceleration of
the Pioneers and other deep space probes, also revealed in 1998
\cite {Anderson1998}
\cite {Turyshev1999}.
Significantly,
the $\Lambda \sim H^2$ formula further yields an exact match,
as I shall show,
within the Friedmann-Robertson-Walker (FRW) formalism, implying that
it is the traditional interpretation of relativity,
not its mathematical framework,
which is fundamentally at fault.
I have also stated that
the effected would resolve hitherto unexplained evidence of
a past expansion of the earth
\cite {Runcorn1965}
\cite {Wesson1973}
\cite {Wesson1999pvt}.

Though the phenomenon is not directly contradicted by quantum mechanics,
its inference from the fundamental relativity of scale
\cite {Prasad2000c}
could not be satisfactory without
a complete understanding of the processes involved,
especially as it attributed an observed redshift $\sim Hr$
solely to the contraction of the observer,
instead of to the incoming light
as in both the standard model and alternative theories.
The requisite knowledge of the quantum absorption process
has now been achieved
\cite {Prasad2000a}
\cite {Prasad2000b},
allowing the inference that the Hubble flow is virtual,
to be fully explained on basis of the contraction
in \S\ref{s:quant}.

Accordingly, I shall first show, in \S\ref{s:mech}, that
the phenomenon,
which is the natural form of plastic flow
to be expected under curved stress,
yields the correct order of magnitude, and then 
describe how it gets amplified by repetitive tidal action.
I shall next derive the two component model
I had previously proposed to explain the Pioneer anomaly
\cite {Prasad1999},
proving, from consideration of
the onboard signal path and the range involved
in the measurements, that
\emph{the anomaly mathematically implies planetary Hubble flow}
(\S\ref{s:signal}), and that
the reported variations of the anomaly
do not permit a different explanation
(\S\ref{s:vars}).
I shall argue that
the phenomenon would operate uniformly at all microscopic scales
because of thermal balance,
for consistency in the quantum picture.
Lastly,
I shall briefly discuss
how it explains the apparent doubling of the earth's radius
indicated by a number of geophysical and palaeological studies
\cite {Runcorn1965}
\cite {Wesson1973}.

\section {Tidal contraction and expansion}
\label {s:mech}

\Fig[th] {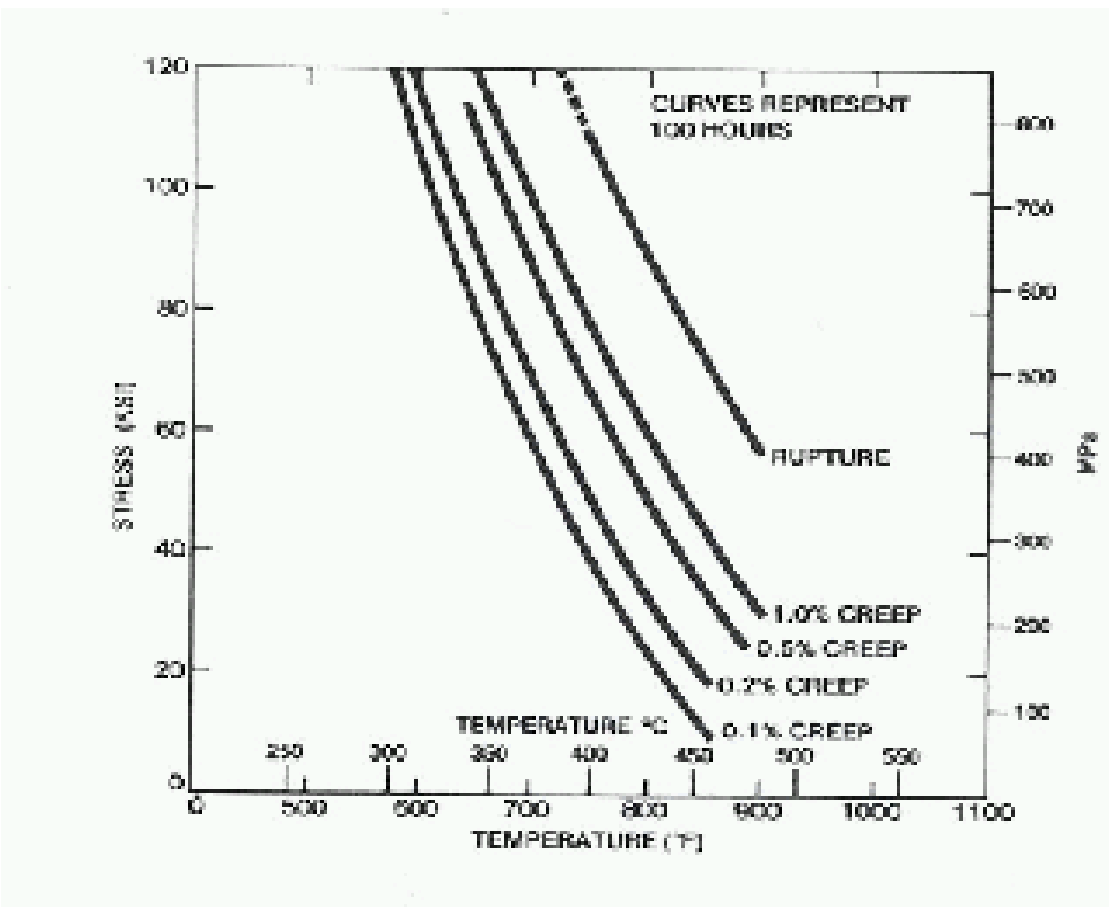}
{Creep curves for a titanium alloy\cite{Titanium}}
{Creep}

Even under earth's gravity,
which is the larger force to be considered in the present context,
the expected plastic deformation rate in any solid material
is far too small to be detectable,
unless accelerated by melting the material.
However,
as previously discussed
\cite {Prasad1999},
we need a deformation rate of only $10^{-19}$-$10^{-18}$~\s
for explaining the variations in the Pioneer anomaly.
The detailed mechanisms of dislocation and plastic flow,
which have been well studied at much higher stresses,
are not necessarily applicable at this scale, and
the detailed behaviour in a specific spacecraft
would be difficult to compute,
if not of questionable relevance,
as substantially the same behaviour is exhibited
by multiple spacecraft
with substantial differences in material and fabrication.
We seek only a general description of
the deformation process, therefore, and proceed from
that of plastic flow under steady stress conditions
\cite [vol.XIV p.37-38] {McGrawEncl},
\begin {equation} \label {e:creep}
	\dot{\epsilon}
	=	\sigma ^n \, e^{- W / k_B T} ,
\end {equation}
where $n$ denotes an empirical constant,
$T$ is the temperature, and $\sigma$, the stress.

The immediate concern is to verify that
eq.\ (\ref{e:creep}) could yield
the desired order of magnitude for the small centrifugal stresses
in our spacecraft,
which is clearly determined by the exponential factor.
Equating $e^{W / k_B T}$ to $10^{-19}$~\s,
we get $W \approx 1$~eV at $300$~K,
which seems quite reasonable. 
For example,
the creep curves in Fig.\ \ref{f:Creep}
indicate $n \approx 0.29$ and $W \approx 1.2$~eV.
Both $n$ and $W$ could vary considerably
not only between materials, but even in a given sample,
so the computation serves only to establish
only the plausibility of this phenomenon as
a cause of the anomaly.
Moreover,
the effective temperature $T$ for the spacecraft is
probably significantly less than $300$~K, and
its exponential contribution needs to be magnified
by tidal action, as explained next.

\Fig {expcon}	{Cumulative dislocations}	{Disloc}

Net expansion should result from the plastic flow in any case,
because the centrifugal force is radial, and
would cause the onboard material to stretch laterally
as it is pulled outward.
A lateral expanding stress is thus created by
the cylindrical curvature of the centrifugal field, and
the resulting dislocations would leave gaps in the lattice,
forming microscopic breakages and causing creep.
On earth,
the spherical geometry of its gravitational pull produces
a lateral compressive stress,
forcing the dislocations to fill interstitial spaces and
squeeze out atoms at the surface.
The deformation rate due to these steady forces
appears to be too low to account for the anomaly,
as will be explained.

Considerable amplification is provided by tidal action,
which not only periodically stretches the lattice,
but rotates in the transverse plane,
causing differential stretching between neighbouring regions of
the lattice.
As illustrated in Fig.\ \ref{f:Disloc},
this significantly raises the probability of dislocation
above that in a uniform stress,
such as provided by the centrifugal and gravitational forces,
where the neighbouring regions
would be stretched or compressed to the same extent.
For example,
atom $b_2$ is subjected to stronger forces
from $a_1$ and less from $a_3$ than it would ordinarily,
so that it could be pushed or pulled out by $a_1$,
depending on the sign of
the simultaneous lateral stress $\sigma$.
The enhanced probability of dislocation means that
the tidal action effectively lowers the dislocation ``barrier'' $W$,
by periodically injecting tidal energy $W_t$ into the lattice,
so that the exponential factor in eq.\ (\ref{e:creep})
becomes $e^{(W_t - W) / k_B T}$.
This is analogous to the action of the gate signal in
a field-effect transistor (FET),
which modulates the depletion region, and
thence the net conductivity in the channel;
the analogy is not perfect, however, because 
the dislocation density quickly reaches thermal equilibrium,
returning the instantaneous plastic deformation rate
to its steady stress value.
However,
the tidally induced dislocations would be of substantial density
because of the periodicity of the lattice, and
the incremental radial creep would make them irreversible
upon withdrawal of the injected energy $W_t$ at each ebb.
The resulting creep rate would therefore
be proportional to the stress times the spin,
which defines frequency of repetition of this tidal process.

Furthermore, in steady plastic flow,
interactions between dislocations lead to
the formation of lattice-like structures.
The nonlinearity represented by $n$ in eq.\ (\ref{e:creep})
occurs because the dislocation lattices change dynamically
with the stress and flow rate.
In the tidally induced flow,
the incremental creeps are purely transient, as described,
and given the slowness of the phenomenon,
the lattice conditions are not significantly altered
between the successive tidal sweeps,
so that the interactions between dislocations during these times
are not much changed in the course of the flow.
As a result,
we would have $n = 1$ identically for this component of the flow,
yielding the formula
\begin {equation} \label {e:lcreep}
	\dot{\epsilon}
	=
		k_\sigma \, \omega^\alpha \, \sigma \, e^{- W / k_B T} ,
\end {equation}
where $k_\sigma$ and $W$ would both be different
for the earth and for each spacecraft, and
$\alpha = 3$ in the case of the spacecrafts
including the $\omega^2$ dependence of the centrifugal force, and
$1$ in the case of the earth.

Since we are only considering gross behaviour,
we cannot depend on $k_\sigma$ and $W$ alone to be responsible
for matching the orders of magnitude
between the two components of the anomaly,
attributed to contraction on earth and expansion onboard,
respectively.
Variations in $T$,
due to varying distance from the sun and
the aging of the RTG (radioactive thermoelectric generator),
would cause the latter component to vary over each mission,
as separately considered in \S\ref{s:vars}.
For the moment,
we still need to verify that
the $\omega^\alpha \sigma$ product is similar between
the earth and the spacecraft, in order that
the creep rates can be of similar magnitudes,
\viz $10^{-19}$-$10^{-18}$~\s.
Taking Galileo as example, we find that
its $2.5$~ton mass would produce a net centrifugal force
of $18$~N at its spin axis,
which is where the telemetry antenna and circuits
are generally housed,
about $150$ times smaller than
the gravitational force on a comparable mass on earth.
However,
this particular spacecraft spins nominally at $3$~rpm,
$3 \times 24 \times 60 = 4320$ times faster than the earth,
which leaves a factor of only $28$ to be made up
by $k_\sigma$, $W$, $T$ and possibly geometrical factors
like $\pi$.

I have thus shown that
the repetitive action of a gravitational tidal force
in the presence of a lateral stress
produces cumulative expansion or contraction,
depending on the sign of the stress, and further that 
the mechanism would be of the right magnitude
for explaining both
the Pioneer anomaly and the Hubble flow.

\section {Signal path analysis}
\label {s:signal}

I now establish,
by analysing the actual process of measurement,
how the Pioneer anomaly indisputably indicates
the presence and involvement of both mechanisms,
of expansion of the spacecraft and of contraction on earth.
Moreover,
the analysis will show that
any complete explanation must introduce
phenomena of identical behaviour, \ie
that a fundamentally different explanation is impossible.

The indication of the anomaly comprises
an almost linearly increasing Doppler residual
in the Pioneer's ranging signal,
which is equivalent to an unmodelled acceleration
acting on the spacecraft
in the same direction as the sun's gravitational field
\cite {Anderson1998}
\cite {Turyshev1999}.
That is,
the data is not of the perceived acceleration itself,
but of an increasing residual $\dot{\omega}$,
so we must examine this first
before turning to relativistic or mechanical causes
for explaining actual acceleration.
The reason such hypotheses by other researchers is that
though the ranging procedure
involves a different downlink frequency from the uplink,
the possibility of drift,
due to heat and circuit degradation, 
was assumed to have been completely eliminated
by the use of a phase-locked loop (PLL)
\cite {Bender1989}
\cite {Vincent1990}
\cite {Anderson1993}.
%\cite {referee2000}.

What has been overlooked is that
the PLL cannot eliminate physical processes
directly affecting the \emph{signal path},
which would however impact the phase $\phi = r_t / \lambda$,
where $r_t$ is the effective signal path and
$\lambda$, the carrier wavelength,
taken, for simplicity of argument,
to be the constant over the entire round trip.
The total path $r_t$ then comprises
linear segments to and from the spacecraft, of length $r$, plus
an onboard segment $r_s$
representing the delay onboard the spacecraft,
as illustrated in Fig.\ \ref{f:Sigpath}.
The Doppler residual of Pioneer 10
\cite {Anderson1998}
is thus indeed an acceleration
\begin {equation} \label {e:accel}
	\dot{\omega} \equiv \ddot{\phi}
	\sim \lambda^{-1} \ddot{r}_t
	\sim \text {constant} ,
\end {equation}
but the second derivative is of $\phi$, not $r_t$ itself,
as required for a kinematic acceleration.
Instead, by using the signal path,
we are no longer concerned with precise distribution of
propagation delays and the refractive index, and
can obtain the net impact on the path length more directly
by integrating this phase acceleration,
which yields
\begin {equation} \label {e:thubble}
\begin {split}
	\int \ddot{\phi} \; dt
	=
	\int \lambda^{-1} \ddot{r}_t \; dt
	=
	\lambda^{-1} \dot{r}_t t
	&\approx
	\text {constant} \times t
\\
\text { or }
	\dot{r}_t = h c t = h r
	,
\end {split}
\end {equation}
where the reported variations in the anomaly
would be contained in $h$.
Eq.\ (\ref{e:thubble}) has the exact form of Hubble's law,
as previously pointed out
\cite {Prasad1999}, and
the reported magnitude of the anomaly,
$2.8 \times 10^{-18}$~\s
\cite {Anderson1998},
makes $h$ approximately equal to the Hubble flow $H$,
whose accepted value
$75~$~km/s-Mpc is equal to $2.43 \times 10^{-18}$~\s.
The Pioneer anomaly is
thus direct evidence of planetary Hubble flow, and
\emph{there is no mathematical room for another explanation}.
I show further, in \S\ref{s:quant}, that
this conclusion is indeed consistent with the FRW model at this range,
and that its existence was unobvious
only because of current prejudices in relativistic cosmology.

\Fig {sigpath}	{Telemetry signal path}		{Sigpath}

The analysis invalidates Anderson \etal's contention that 
a comparable effect is absent in planetary ranging data,
as they considered only Mars and Venus,
both less than $0.75$~AU from earth.
As I have already shown
\cite {Prasad2000c}
\cite {Prasad1999},
the ranging imprecision,
which itself follows Hubble's law in being proportional to $r$,
is at least 10 times too coarse for
the detection of planetary Hubble flow
indicated by the anomaly.

We can distinguish two separate contributions to $h$, however,
physical expansion of spacecraft material,
causing the onboard path segment $r_s$ to increase,
yielding $\dot{r}_s \sim h_s$, and
expansion of the uplink and downlink spatial segments,
each of length $r$,
the latter in effect describing expansion of
space instead of matter.
The latter notion would lead to contradiction when
applied to our terrestrial unit referents of scale
if interpreted according to the prevailing semantics
of relativity
\cite [p.719] {MTW}
\cite [p.197] {Rindler},
but it has a precise formal interpretation
in terms of the relativity of scale
\cite {Prasad2000c}, and
the likelihood of the requisite contraction
at this rate has just been established in \S\ref{s:mech}.
We may accordingly break up $h$ as
\begin {equation} \label {e:breakup}
	\dot{r}_t	
	= h c t
	= h_c c \, t_{2r} + h_s c \, t_{s}
	= 2 h_c r + h_s r_s
	,
\end {equation}
where $h_c$,
to be attributed to contraction on earth, 
would be the constant component in all six missions.
It would be recognised that
$h_c$ would also contribute to the onboard segment,
but this contribution can be conceptually absorbed in 
the spatial segments,
leaving the onboard material expansion as
the only contribution in $h_s$;
in any case, as $r_s \ll r$ spatially,
the net $h_c$ contribution from the $r_s$ would be negligibly small.
Conversely implied is that
the $h_s$ contribution must be disproportionately larger
than the spatial extent of $r_s$,
which could be due to $k_\sigma$ being much larger
than for earth-bound materials, but this seems unlikely,
as the ratio $r/r_s = O(10^{12})$ is rather large.
On the other hand,
substantial phase delay occurs over $r_s$, and
would be proportionally increased by
physical expansion of the circuits,
which has already been shown, in \S\ref{s:mech},
to be not insubstantial.

\section {Correlation of variations}
\label {s:vars}

While I have established planetary Hubble flow as
the only possible explanation for the principal component $h_c$,
I have yet to show that
material expansion onboard is the only one for $h_s$.
I do this now by deriving,
from the notions of \S\ref{s:mech},
the empirical model I had previously proposed
to account for all variations of the anomaly
\cite {Prasad1999},
which are particularly described by the best-fit curve given by
Turyshev \etal
\cite {Turyshev1999}, \viz
\begin {equation} \label {e:spin}
	h_s =
		k_s (\theta) \;
		\omega^\alpha g^\beta \;
			| \bold{\hat{\omega} \wedge \hat{g}} |
	,
\end {equation}
where
$\omega$ denoted the spacecraft's spin,
$\bold{g}$, the net gravitational force acting on the spacecraft,
mostly due to the sun.
The anisotropy factor $k_s(\theta)$ means that the expansion,
per each tidal sweep,
cannot be assumed to be the same in all directions,
as the spacecraft are not homogeneous blobs of matter.
Given the nominal spin of $3$~rpm,
the $20$~Hz peak fluctuation rate during Galileo's earth fly-by
\cite {Turyshev1999},
seems to be a clear symptom of
physical features in the onboard signal path
individually subtending
$360/(20 \text{ s/cycle} \times 20 \text{ Hz})
= 0.9$\degree{} at the axis,
which seems quite reasonable judging from
the diagrams available on NASA's Galileo Web site.

It was subsequently pointed out to me that
%\cite {referee1999} that 
this inclusion of $\bold{g}$ cannot be precise,
as tidal action depends not on $\bold{g}$,
but on its gradient $\nabla \bold{g}$,
as implicitly considered in \S\ref{s:mech}.
Eq.\ (\ref{e:spin}) was, however, intended to be capable of
accommodating any order of derivative whatsoever
that might be uncovered by subsequent investigation;
for example, since
\begin {equation} \label {e:tidalforce}
	|\nabla \bold{g}|
		= \frac{\partial}{\partial r} \frac{-G M_\odot}{r^2}
		= \frac{2 G M_\odot}{r^3}
		= \frac{2 g}{r}
		= \frac{2 g^{3/2}}{\sqrt{G M_\odot}}
	,
\end {equation}
a direct dependence on $\nabla \bold{g}$
would have been represented by $\beta = 3/2$.
However,
as the magnitude of $\nabla \bold{g}$ even at $1$~AU
is only $7.93 \times 10^{-14}$~N/kg-m,
tidal action by itself could not have been the driving force
for the onboard expansion,
but the centrifugal force, of the order of $18$~N, clearly could,
for which $\beta = 0$.
The angular dependence, denoted by the vector product, survives,
however, as the tidal action ``gates'' the expansion,
as described.
We must also include
the temperature dependence from eq.\ (\ref{e:lcreep}),
which was unfortunately omitted in eq.\ (\ref{e:spin}),
and absorb $k_\sigma$ into $k_s$,
to arrive at the correct model
for the effect of onboard tidal distension,
\begin {equation} \label {e:spin2}
	h_s =
		k_s (\theta)
		\;
		\omega^\alpha
		\;
		| \sin ( \hat{\omega}, \hat{g} ) |
		\;
		e^{-W / k_B T }
	.
\end {equation}

As described in 
\cite {Prasad1999},
the angular dependence, due to the fact that
the spin axis generally points toward the earth and
subtends an angle $\angle (\hat{\omega}, \hat{g})$
with the sun's gravitational pull, 
would explain the almost linear falloff of the anomaly
between $5$ and $40$~AU.
and not the magnitude of $\bold{g}$ is material,
explains why the linearity holds in the $5$-$40$~AU range
regardless of the $r^{-2}$ falloff of the sun's pull and
the $r^{-3}$ falloff of its tidal action.
At the other extreme,
both at the perihelion and during earth-flybys,
the spin axis would be normal to the sun
at least at the instants of observation,
in order to ensure the maximum angular separation
from the sun for the ground telemetry antennas;
this, together with the generally higher temperature
due to proximity to the sun,
seems adequate to account for
very high values of the anomaly seen at these times.

The angular dependence had also prompted a conjecture that
the residual difference between the Pioneers 10 and 11
could be at least partly due to
the angles made with respect to the galactic centre.
A more precise argument can now be made in retrospect,
that though the sun's gravitation even at $70$~AU is much stronger
than the galactic field,
$1.21 \times 10^{-6}$~m/s$^2$
\vs
$1.9 \times 10^{-10}$~m/s$^2$,
the solar contribution to the onboard expansion
would be vastly diminished,
as the spacecraft spin axes would be pointing
almost directly toward the sun.
Eq.\ (\ref{e:spin2}) now provides a second factor
that could be just as significant, \viz the temperature $T$,
because Pioneer 11,
which exhibits the larger residual anomaly,
is both slightly younger,
so that its RTG generates more heat than Pioneer 10's, and
is headed into the heliopause.
However,
differences in the construction of these craft,
affecting $k_\sigma$ and $W$, as already explained,
may turn out to be more influential than
either of these hypotheses.

My contention that
this is the only possible explanation of
the variations in the anomaly, appears to be no weaker than
the mathematical inference of planetary Hubble flow
given above (eq.\ \ref{e:thubble}).
The oscillatory characteristic of the NASA-JPL best-fit curve
cannot be modelled without introducing a sinusoidal factor.
Its apparent synchronisation with the earth's orbit and
asymmetries consistent with occlusion by the sun
\cite {Prasad1999},
suffice to relate its phase to the earth's relative position;
as the linear falloff from $5$ to $40$~AU
is then adequately modelled by the sine factor,
we cannot have any other
that would change substantially in this range.
Since an oscillation ($20$~Hz)
substantially greater than the spin frequency ($1$-$3$~rpm)
was observed,
we must include an anistropy factor $k_s (\theta)$.
As we have no room left for  $\bold{g}$ or its derivatives,
we need an amplification factor that 
particularly contributes at the perihelions, where both
the tidal action $\nabla \bold{g}$ and the temperature $T$
might be substantial,
but at the other extreme, only $T$ would survive to contribute
to the residual difference between the two Pioneers.
The Boltzmann factor $e^{-W / k_B T}$ appears to be 
the simplest form for the inclusion of $T$, and
the scale factor $W$ then
acquires natural interpretation as activation energy.
We also find,
from eqs.\ (\ref{e:thubble}) and (\ref{e:breakup}),
that the result must represent
an expansion of the signal path in some way,
suggesting plastic flow, and discover not only that
the Boltzmann form indeed occurs in
the theory of dislocations and creep,
but that the anomaly is consistent with
the known values of creep under macroscopic stresses.
Finally,
the dependence on the subtended angle
indicates an involvement of tidal action,
but this cannot by itself provide net flow;
a driving stress is required and is readily identified
with the centrifugal and gravitational forces
on the spacecraft and on earth, respectively.
We still need a dependence on spin,
not only in order to match the magnitudes of
the $h_c$ and $h_s$ components,
as explained in \S\ref{s:mech},
but to also account for the changes in the anomaly
coinciding with the changes in the spin frequency
\cite {Turyshev1999}.

\section {Time-variant quantum scale}
\label {s:quant}

It should be clear that
the mechanism of tidal contraction or expansion
is a general principle
that would hold for any particulate structure of matter
in which the particles have well-defined mean locations,
as there would be attractive and repulsive interactions
between particles at these positions
to which the reasoning of Fig.\ \ref{f:Disloc} could be applied.
We would expect to find it operating in glasses, for example, and
within Bose condensates not necessarily made of atoms,
but not within liquids or gases,
which has a geophysical significance
to be described later.
There is also disparity in the rates of contraction
one would expect between materials,
as the rates depend on $k_\sigma$ and $W$.
The disparities are of interest
as the particulate density would not be preserved
in the interior of the lattice, and
could become detectable via their impact on
the electronic energy levels, for instance.
But this appears unlikely,
as the thermal interactions would be sufficient,
given the slowness of the phenomenon,
to even out the rate disparities and
make the contraction uniform and continuous on earth.
This is particularly true in the measurement of spectra,
which by definition require the observations to last long enough
for the instrument state to settle.

A more important consequence
is the appearance of a Hubble flow in an otherwise steady universe
to the terrestrial, shrinking observer
\cite {Prasad2000c},
which can now be fully appreciated in terms of the quantum picture,
as follows.
Recall that in every quantum measurement,
the outcome is determined by an amplitude of
the form $\iprod{\psi}{\phi}$,
where if $\ket{\phi}$ denote the variable being measured,
$\bra{\psi}$ must represent the data value that would be returned
whenever the variable would be subsequently left
in the state $\ket{\psi}$,
which, of course, is quantified by the probability 
$|\iprod{\psi}{\phi}|^2$.
As a result,
$\bra{\psi}$ must represent
a macroscopic physical state of the observing system, and
this makes every distance-related measurement
susceptible to the effects of the contraction,
as $\bra{\psi}$ itself becomes time-variant, causing
the amplitudes $\iprod{\psi}{\phi}$
to vanish except for $\ket{\phi}$s
which are time-varying the same way.

\Fig {doppler}	{The Parker eigenstates}	{Doppler}

Fig.\ \ref{f:Doppler} shows
a wave being received by a detector, and
the latter's ongoing shrinkage, indicated by the arrow.
Since a photon absorption must correspond to 
the smallest possible change between
the stationary states of the detector,
which means a whole ``antinodal lobe'',
\ie the portion of a standing wave between adjacent nodes, 
the detector can only detect waves that
fill an exact number of antinodal lobes in the detector,
as depicted for the lower order wave in the figure.
The properties that
make these antinodal lobes specially significant are:
\LN
\def\theenumi{\roman{enumi}}
\item \label {i:partition}
	Their energies are dependent only on
	their electromagnetic amplitudes, and
	not \apriori on the frequency or wavelength,
	which makes them the right classical candidates,
	in place of whole modes,
	for thermal equipartition.

\item \label {i:lobes}
	A transition between stationary modes
	necessarily involves
	an integral change in the number of antinodal lobes.
	Together with (\ref{i:partition}),
	this yields Planck's law, $E = h \nu$.\footnotemark[1]
\footnotetext[1]{Not to be confused
	with the $h$ in eq.\ (\ref{e:thubble}).}

\item \label {i:cavity}
	An equilibrial equipartition over such lobes
	does yield Planck's law for the cavity spectrum
	\cite {Prasad2000a}, and
	these two principles of
	stationarity and antinodal equipartition
	have been further shown to be sufficient
	for deducing the correct form of
	quantisation in the interactions of matter
	\cite {Prasad2000b}.
\NE
Properties (\ref{i:lobes}) and (\ref{i:cavity}) are the reason that 
every photon detector behaves as a resonant cavity;
property (\ref{i:partition}) of course provides
the crucial conceptual link to classical mechanics
that has been missing in the 20th-century physics,
which enables us to consider the physics of shrinking observers
beyond simplistic philosophical terms
(\cf \cite [p719] {MTW}), and
to deduce a redshift conforming to Hubble's law, as follows.

To obtain the form of the stationary states of a shrinking detector,
we take the stationary waves
$e^{i(\bold{k}\cdot \bold{x} - \omega t)}$,
for the nonshrinking detector, and
insert an increasing scale factor $a(t)$ to compensate
for the shrinkage along the linear dimension.
This yields
$e^{i(\bold{k}\cdot [a \bold{x}] - \omega t)}$,
which is Parker's solution for the wave equation in the FRW metric
\cite {Parker1988}
\begin {equation} \label {e:FRW}
	ds^2 = - dt^2 + a^2 (t) (dx^2 + dy^2 + dz^2)
	,
\end {equation}
which we in turn interpret, in our context,
as describing the \emph{apparent} scaling of all space
in the perspective of the shrinking observer,
to whom the detector eigenstate must appear to be a uniform wave.
In reality,
as would be seen by nonshrinking observers,
the wavelength of this eigenstate must correspondingly decrease
with distance as shown.
The shrinking detector can thus strike a momentary resonance,
as required for photon detection by (\ref{i:lobes}),
at wavelength $\lambda_b$ at range $r_b$ and
at $\lambda_c$ at range $r_c$,
both differing from the wavelength $\lambda_a$ at the detector.
This apparent redshift, seen by the shrinking observer,
may be thought of as a virtual Doppler effect,
as the apparent expansion of space, eq.\ (\ref{e:FRW}),
implies a virtual recession of all objects.
However,
to explain the effect from
the perspective of the nonshrinking observers,
we cannot simply count the wavefronts crossing
the detector boundary ($r_a$) as in prior theory,
because that would fail to take the contraction into account and
thus include waves which cannot be detected.

We are therefore forced to count only the waves
which would measure a fixed number of antinodal lobes
within the detector,
which would be of
$\lambda_b$ if they had started out at $r_b$ and
$\lambda_c$ if they had begun at $r_c$,
because the detector would have shrunk by
$r_a/r_b = \lambda_b/\lambda_a$ and
$r_a/r_c = \lambda_c/\lambda_a$
over their respective times of flight $r_b/c$ and $r_c/c$.
Additionally,
since we clearly have no logical room for issues of dispersion,
as in Parker's theory,
$a(t)$ must be strictly linear,
\begin {equation} \label {e:ahubble}
	a (t)
	=
		\dot{a} t
	\equiv
		(\dot{a}/a) \, t
	=
		H_r t
	,
\end {equation}
and
$\lambda_a : \lambda_b : \lambda_c = r_a^{-1} : r_b^{-1} : r_c^{-1}$,
literally following Hubble's law.
This is readily interpreted as saying that
only the instantaneous value of $H_r$, the contraction rate,
can possibly affect our immediate observation.

More particularly, eq.\ (\ref{e:ahubble}) yields
the acceleration formula $\ddot{r} = H_r^2 r$,
which, as stated at the outset,
exactly matches the observed $\Lambda$
\cite {Prasad2000c}.
The linearity clearly holds at any scale of measurement
in our theory,
unlike the current thinking in relativistic cosmology,
where both $H$ and $\Lambda$ are commonly assumed to operate
only on intergalactic scale.
In fact,
the equations of the relativistic theory
(\cf \cite [p98] {Wald})
do not involve any variable or relation
to formally incorporate this prejudice, and
it is therefore not surprising that
they do predict an incremental acceleration
\cite {Cooperstock1998}
\begin {equation} \label {e:cooperstock}
	\ddot{r}
	=
		- \frac{4 r}{3 t^2 \omega^2}
	=
		- 3.17 \times 10^{-47}
		\text{ m/s$^2$ for earth}
	,
\end {equation}
where $\omega$ is the orbital frequency ($2 \times 10^{-7}$~\s) and
$t$, the present epoch ($6.3 \times 10^{17}$~s).
This at first seems to be too small
to correlate with the Pioneer anomaly,
but the involvement of $\omega$ obfuscates the scalability
of eq.\ (\ref{e:cooperstock}).
Instead,
from our preceding formula for acceleration,
which in effect includes $\Lambda$, we directly obtain
\begin {equation} \label {e:match}
	H_r
	=
		\sqrt {- \ddot{r} / r}
	=
		2.178 \times 10^{-18}
		\text { s$^{-1}$}
	,
\end {equation}
closely matching the anomalous time dilation rate
$2.8 \times 10^{-18}$~\s
\cite {Anderson1998},
and supporting all earlier conjectures of
the cosmological connection of the anomaly
\cite {Rosales1998}
\cite {Ellman1998}
\cite {Prasad1999}.

More importantly, it means that
the conceptual foundations of relativistic cosmology,
based on Mach's philosophy
\cite {Einstein1911},
are not only imprecise,
in not taking the observer's referents as the basis of relativity
\cite {Prasad2000c},
but also at variance with the empirical evidence
of both $\Lambda$,
for which the standard model ideas lead to speculations of
large scale repulsion, and
of the Pioneer anomaly,
for which modifications to gravitational field theory
were being considered
(\cf \cite {Anderson1998}).
It also invalidates
the big bang and related notions of the standard model,
though it may yet be possible to reinterpret some of its results
in light of the present discovery of tidal contraction,
as I did for the Doppler theory of the Hubble redshift.
The $t$ in eq.\ (\ref{e:match}) cannot, of course,
be interpreted as the present age of the universe
in the present theory, and
is recognisable as $H$ in disguise,
obtained from the large scale measurements.

\section* {Conclusion}

It should be mentioned that
any ongoing contraction could reproduce the Hubble flow,
including descent within a gravitational well;
but the descent would have to be at over $128$~km/s
in order to explain the Hubble flow
\cite {Prasad2000c}.
The Pioneer anomaly is
the first direct evidence of planetary Hubble flow
resulting from a contraction of referents, but as just shown,
our prejudicial views prevented us
both from recognising this and
from predicting it in prior theory.
This is just as well,
because we might then have lacked the motivation
to explain the variational part $h_s$ (\S\ref{s:vars}), and
the tidal mechanisms would have been much harder to uncover.
Instead,
as even this connection was not obvious at the outset,
precise logical foundations of both relativity and quantum mechanics
have been discovered in the course of this investigation,
while the key mechanism itself has been demonstrated to be
of macroscopic origin and to scale correctly
with routine measurements of creep.

While the problem of the Pioneer anomaly
has been completely solved thus essentially by creep,
the significance of the result is hard to overestimate.
As formally predicted and now shown,
it invalidates \emph{all} our current notions of cosmology,
as the Hubble redshift,
on which they were fundamentally based,
has been shown to be the result of
a strictly terrestrial mechanism.
It is worth noting that
the arguments of \S\ref{s:quant} are not dependent on
the acceptability of the logical foundations mentioned above,
as the only property of antinodal lobes critically used,
(\ref{i:lobes}),
follows immediately from Planck's law $E = h \nu$.
The uniformity of contraction was independently argued
to result from a basic rule of spectral measurement,
that sufficient time be allowed beyond
that required for thermal stability of the detector's state,
which is consistent with Landauer's principle
relating thermalisation and data states
\cite {Landauer1961}.
While the contraction itself is consistent with ordinary creep data 
in terms of magnitude,
the driving mechanism for the compaction,
or distension aboard the deep space probes,
appears to be sound and adequately scales between
the earth and the spinning spacecraft.
I have also previously shown
that considerable evidence of past expansion of the earth
\cite {Runcorn1965}
\cite {Wesson1973},
which remain unresolved
\cite {Wesson1999pvt},
as well as the measured lunar recession
\cite {Lunar1994},
are consistent with the Hubble flow on these scales
\cite {MacDougall1963}
\cite {Prasad1999},
and this, as shown in \S\ref{s:quant},
would be consistent with the relativistic theory,
but not with the prior views.
Against this,
I make no attempt to reinterpret or dismiss
the existing results of the big bang theory, such as
the cosmic microwave background (CMB), Olbers' paradox or inflation,
as these issues appear to be secondary to
that of correctness and acceptability of the terrestrial contraction.

The explanation of the past expansion evidence 
is interesting in its own right,
as we no longer have to deal with an overall expansion of the earth. 
This would have required
difficult hypotheses of ongoing creation of matter,
which fell short on the required magnitude by a full order
\cite {Wesson1973}
\cite {NarliKem1988}.
The evidence is principally that
the continents must have not only once formed one contiguous mass,
but also that this mass should have once completely covered the earth
\cite {Runcorn1965}.
The present theory resolves this difficulty perfectly,
as the contraction would have caused the sialic masses
to break apart, forming the tectonic plates, and
to continuously widen the gaps between the continents,
independently of the tectonic motions,
as the contraction does not apply to liquid or gaseous matter.
The total widening should be about $3.77$~mm/y,
corresponding to an apparent expansion of the earth's radius,
as seen by earth-bound observers
to whom the continents would not appear to be shrinking,
by about $H \times 6400 \text{ km} = 0.6$~mm/y
\cite {Creer1965};
this may be verifiable by GPS measurements now or in the near future.

It was also of concern to me that
attributing a third of the lunar recession to the Hubble flow
\cite {Prasad2000c}
\cite {Prasad1999}
might lead to inconsistency with the known cause of tidal friction,
which is necessary to explain the slowing down of the earth's spin.
This problem too is now resolved,
as we can now recognise, from \S\ref{s:mech}, that
the Hubble component itself requires the earth's spin
to do work in shrinking the sialic matter.

We would also expect find this mechanism operating
on every planetary body with a solid surface and
subject to tidal action.
We do find such markings on Europa,
which is subject to strong tidal forces from Jupiter and Ganymede,
and on Mars,
which does suffer tidal action from the sun,
but the contribution remains to be estimated
in either of these cases.
Correspondingly,
we expect to find an expansion occurring
in every spinning spacecraft, but none in the others;
since spin seems to be necessary for achieving the stability needed  
for the measurement
\cite {Anderson1998},
the antennae and electronics subsystems used for the ranging
would have to be despun for verifying this conclusion.

\acknowledgements

Special thanks are owed to
Bruce G Elmegreen and a referee
for posing the key issues that led to this answer, and to
ASM International for their kind permission to reproduce
the creep curves in Fig.\ \ref{f:Creep}.

% end
% --- figs.tex ---

%\Fig{anom}	{Geometry of the anomaly}		{Anom}
%\Fig{struct}	{Spacecraft construction schematic}	{Struct}
%\Fig{orbit}	{Orbital geometry}			{Orbit}
%\Fig{galactic}	{Galactic contribution geometry}	{Galactic}

% === cat *.bbl ===

\end {document}